\documentclass[printer]{aa}
\usepackage{amsmath}
\usepackage{epsfig}
\usepackage{graphicx}
\usepackage{rotate}
\def\INTEGRAL{{\it INTEGRAL }}
\def\Ginga{{\it Ginga }}
\def\ASCA{{\it ASCA }}
\def\grsim{\stackrel{>}{\sim}}
\begin{document}

\sloppypar

%
   \title{INTEGRAL observations of SS433: Results of a coordinated campaign}

   \author{A.M. Cherepashchuk\inst{1}, R.A. Sunyaev\inst{2,3}, 
   S.N. Fabrika \inst{4}, K.A. Postnov \inst{1,10}, S.V. Molkov \inst{2}, 
   E.A. Barsukova \inst{4}, E.A. Antokhina \inst{1}, 
   T.R. Irsmambetova \inst{1}, I.E. Panchenko \inst{1}, E.V. Seifina\inst{1}, 
   N.I. Shakura \inst{1}, A.N. Timokhin \inst{1}, 
   I.F. Bikmaev \inst{5}, N.A. Sakhibullin \inst{5}, 
   Z. Aslan \inst{6,7}, I. Khamitov \inst{6}, A.G. Pramsky \inst{4},  
   O. Sholukhova \inst{4},
   Yu.N. Gnedin \inst{8}, A.A. Arkharov\inst{8}, 
   V.M. Larionov \inst{9}}

   \offprints{cher@sai.msu.ru}

   \institute{Sternberg Astronomical Institute, Universitetsky pr. 13, 119992, Moscow, Russia
	  \and   
              Space Research Institute, Russian Academy of Sciences,
              Profsoyuznaya 84/32, 117810 Moscow, Russia
          \and 
              Max-Planck-Institute f\"ur Astrophysik,
              Karl-Schwarzschild-Str. 1, D-85740 Garching bei M\"unchen,
              Germany,
         \and   
              Special Astrophysical Observatory, Nizhnij Arkhyz, 
              Karachaevo-Cherkesiya, 369167,  Russia
         \and     
              Kazan State University, Kremlevskaya
            str.18, 420008, Kazan, Russia
	\and 
	      TUBITAK National Observatory, Akdeniz Universitesi
              Yerleskesi, 07058 Antalya, Turkey
	\and
		Akdeniz University, Physics Department, 07058 Antalya,
		Turkey
	\and 
	      Pulkovo Observatory, St.-Petersburg, Russia
        \and
              St.-Petersburg University, Russia
        \and  
              University of Oulu, Finland
            }
  \date{}

        \authorrunning{Cherepashchuk et al.}
       \titlerunning{INTEGRAL observations of SS433}
 
   \abstract{
Results of simultaneous \INTEGRAL and optical observations of the galactic
microquasar SS433 in May 2003 and \INTEGRAL/RXTE observations in March 2004
are presented. Persistent precessional variability with a maximum to minimum
uneclipsed hard X-ray flux ratio of $\sim 4$ is discovered. The 18-60 keV
X-ray eclipse is found to be in phase with optical and near infrared
eclipses. The orbital eclipse observed by \INTEGRAL in May 2003 is at least
two times deeper and apparently wider than in the soft X-ray band. The broadband
2-100 keV X-ray spectrum simultaneously detected by RXTE/\INTEGRAL in March
2004 can be explained by bremsstrahlung emission from optically thin thermal
plasma with $kT\sim 30$ keV. Optical spectroscopy with the 6-m SAO BTA
telescope confirmed the optical companion to be an A5-A7 supergiant. For the
first time, spectorscopic indications of a strong heating effect in the
optical star atmosphere are found. The measurements of absorption lines
which are presumably formed on the non-illuminated side of the supergiant
yield its radial velocity semi-amplitude
$K_v=132\pm 9$ km/s.   
The analysis of the observed hard X-ray light curve and the eclipse
duration, combined with the spectroscopically determined optical star 
radial velocity corrected for the strong heating effect,
allows us to model SS433 as a
massive X-ray binary. Assuming that the hard X-ray source in SS433 
is eclipsed by the donor star that exactly fills its Roche lobe, 
the masses of the optical and compact components in SS433 are suggested
 to be 
$M_v\approx 30 M_\odot$ and $M_x\approx 9
M_\odot$, respectively. This provides further evidence that SS433 is a
massive binary system with supercritical accretion onto a black
hole.

   }

   \maketitle

%

\section{Introduction}

SS433 is a massive eclipsing X-ray binary system at an advanced evolutionary
stage. It is recognized as a supercritically accreting microquasar with a
precessing accretion disk and mildly relativistic ($v\approx 0.26\,c$) jets.
Since its discovery in 1978 (Clark and Murdin 1978, Margon et al. 1979),
this unique X-ray binary has been deeply investigated in optical, radio, and
X-rays (see reviews by Margon 1984, Cherepashchuk 1988, 2002 and Fabrika
2004 for more detail and references).

Among a dozen known microquasars, 
SS433 is distinguished by its unique properties.

1) In this system the optical star fills  its Roche lobe and mass
transfer occurs on thermal time scale with the huge rate of $\dot M \sim
10^{-4} M_\odot$ per year and accretion onto the relativistic object is
supercritical (Shakura and Sunyaev 1973).

2) Two strongly collimated ($\sim 1^\circ$) oppositely directed
jets emanate with a velocity of $80,000\pm 1,000$ km per second from the center
of the accretion disk. The jets are observed not sporadically, as in most
microquasars, but persist over tens of years with virtually constant
velocity. 

3) The accretion disk and relativistic jets regularly precess with a period
of 162.5 days. Both the precession period and the disk inclination angle to
the orbital plane ($\sim 20^\circ$) on average remain constant over tens of
years.

4) The source exhibits orbital eclipsing 
periodicity with period $P_{orb}=13^d. 082$. The shape of
the optical light curve varies significantly with precessional
phase (Goranskij et al. 1998ab, Cherepashchuk and Yarikov 1991). The orbital
period is found to be very stable over more than 25 years, 
despite the high mass exchange rate
between the binary components and strong wind mass loss ($v\approx 2000$
km/s, $\dot M\simeq 10^{-4} M_\odot$/yr) from the 
supercritical accretion disk. 

The unsolved puzzles of SS433 still remaining to be solved include: 
(1) the nature of the
relativistic object, (2) the mechanism of collimation and acceleration
of matter in jets to the relativistic velocity $\sim 80,000$ km/s, (3) the
nature of the precessional phenomena in this X-ray binary system
(Chakrabarti, 2002).
  
The  \INTEGRAL observations of SS433 in 2003 discovered a hard (up to 100 keV)
X-ray spectrum in this supercritically accreting microquasar (Cherepashchuk
et al. 2003, 2004), suggesting the presence 
of an extended hot (with a temperature
up to $10^8$ K) region in the central parts of the accretion disk. These new
data made it possible to compare the eclipse characteristics of SS433 at
different energies: soft X-rays (2-10 keV, the \ASCA
data), soft and medium X-ray (1-27 keV, the \Ginga data), hard 
X-rays (20-70 keV, the \INTEGRAL data), and the optical. This comparison
allows us to investigate the innermost structure of the supercritical accretion
disk 
and to constrain the basic parameters of the binary system . 

This paper is organized as follows: Section 2 lists the participants of the
observational campaign. Section 3 describes the X-ray spectra and light
curves obtained. Section 4 presents the analysis of hard X-ray precessional
and orbital variability of SS433. Section 5 reports the results of optical
and infrared photometry. Section 6 describes in detail the simultaneous
optical spectroscopy of SS433 obtained with the 6-m BTA telescope and the
measurements of the optical radial velocity curve. The implications of our
observations for binary masses in SS433 are summarized in Section 7, and
Sections 8 and 9 give the discussion and conclusions, respectively.

\section{Observational campaign}

The coordinated multiwavelength observational campaign of SS433 was 
organized during the  \INTEGRAL observations of the SS433 field in 
March-June 2003 with the participation of the following research  
teams.

1. Special Astrophysical Observatory of the Russian Academy of Science
(SAO RAS). High signal-to-noise optical 
spectroscopy of SS433 at the 6-m telescope.

2. Crimean Station of the Sternberg Astronomical Institute. 
BVR-photometry at the 0.6-m telescope.

3. Kazan State University. Optical photometry
at the Russian-Turkish  
1.5-m telescope RTT-150 at the TUBITAK National Observatory (Turkey).

4. Infrared K-photometry was performed with the 1.1-m AZT-24 
telescope of the Pulkovo
Observatory in Italy, Osservatorio di Campo Imperatore. 

5. Radio monitoring at cm wavelength: the 
RATAN-600 radio telescope of SAO RAS.
The source was found to be in its non-flaring (quiet) state in May 2003 
(Trushkin 2003). 

SS433 was within the field of view of the IBIS 
telescope during the \INTEGRAL observations (March 2004) 
of the Sagittarius arm tangent
region centered on $l=+40^\circ, b=0^\circ$. We included part of these observations
in our analysis. In contrast to the observations in May 2003, 
radio monitoring showed that SS433 had been in its flaring (active) state
(Trushkin 2004).

\begin{figure}
\centering
\epsfig{file=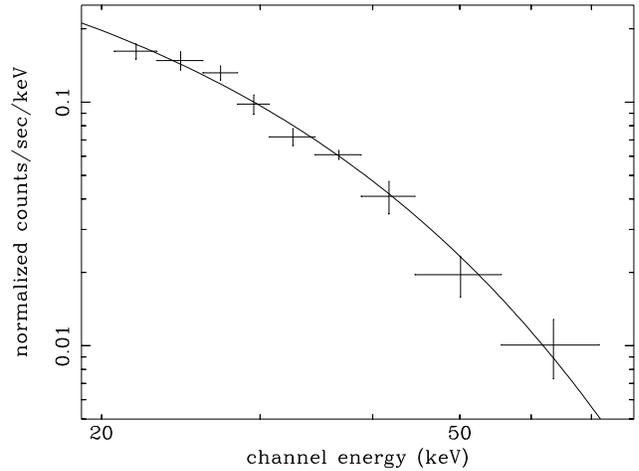,width=0.7\linewidth,angle=-90}
\caption{IBIS/ISGRI spectrum of SS433 collected over \INTEGRAL
orbits 67-69 in May 2003. 
The best fit (solid line) is for an exponential
cut-off with $E_c=14\pm 2$ keV. Reduced $\chi^2=0.32$ for 7 degrees of freedom (dof), 
$CL\approx 95\%$.
\label{fig:IBIS_sp}}
\end{figure}

\begin{figure}
\centering
\epsfig{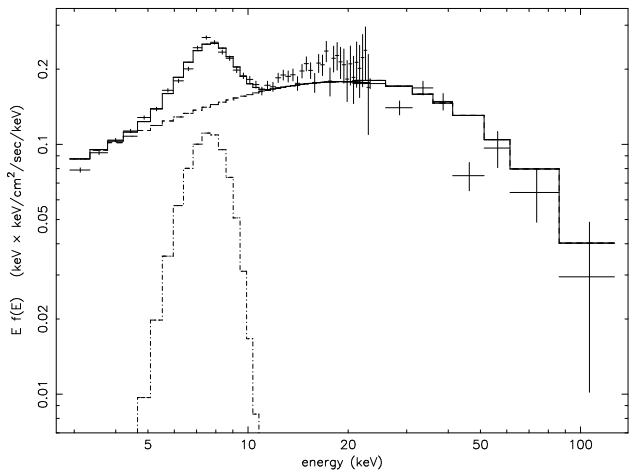}
\caption{RXTE PCA and IBIS/ISGRI spectrum of SS433
obtained during simultaneous {\it RXTE/\INTEGRAL} observations
of SS433 in March 24-27, 2004.
The best fit (solid line) is for bremsstrahlung emission from optically 
thin plasma with $kT=29.0\pm 0.7 keV$ (model "bremss" from the XSPEC package).  
Reduced $\chi^2=0.77$ for 50 dof, $CL\approx 88\%$.
\label{fig:RXTEIBIS_sp}}
\end{figure}

\begin{figure}
\centering
\epsfclipon
\epsfig{file=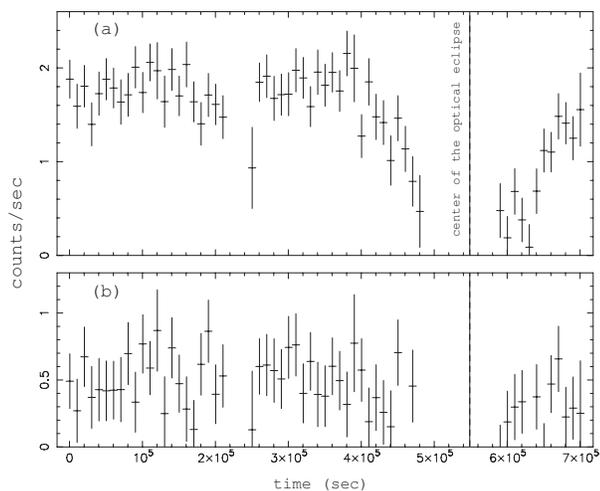,width=0.9\linewidth,angle=0}
\caption{IBIS/ISGRI 20-40 keV (upper panel) and 40-70 keV
(bottom panel) count rates of SS433 (without background subtraction)
obtained in May 2003. The time origin is MJD$=52763.95$.
The egress part of the X-ray eclipse was kindly provided by 
Dr. Diana Hannikainen.  
\label{fig:IBIS_lc}}
\end{figure}

\section{IBIS/ISGRI light curves and spectra}

The 
\INTEGRAL IBIS/ISGRI data were analyzed using  
both publicly available ISDC software (OSA-3 version) and
the original software elaborated by the IKI \INTEGRAL team (see Revnivtsev et al.
2004b for more detail). The latter allows better sensitivity in
constructing maps and spectra.    
The main results can be summarized as follows.

\begin{table}
\caption{Best-fit parameters of the joint RXTE+\INTEGRAL 
spectrum of SS433
observed in March 24-27, 2004. The model is 
bremsstrahlung emission and a Gaussian emission line (from 
the XSPEC package). The energy of the emission line was a free parameter. 
The physical nature of the line is not discussed in the present paper.
}
\begin{center}
\begin{tabular}{lccc}
\hline\hline
 Parameter &  $kT$   & Emission line &  Line \\
           &         & energy        & width  \\
	   &   (kev) &     (keV)      &     (keV) \\ 
\hline
Value       & $28.9\pm 0.7$ &  $7.26\pm 0.03$   &  $1.2\pm 0.04$ \\
\hline 
\end{tabular}
\end{center}
\label{sp_param}
\end{table}

A. The IBIS/ISGRI spectrum of SS433 (25-100 keV) in May 2003
as processed by the OSA-3 software  
can be fitted by a
power-law $\sim E^{-\alpha}$ with photon index $\alpha\approx 2.7\pm 0.13$,
taking into account the 10\% systematic errors. A satisfactory fit is also 
obtained using an exponential cut-off $\exp (-E/E_c)$ 
with $E_c\sim 12-16$ keV 
(Fig. \ref{fig:IBIS_sp}). 

The integrated hard X-ray luminosity in May 2003 was  
$L_x(18-60 \hbox{keV})\sim 4\times 10^{35}$
erg/s, $L_x(60-120 \hbox{keV})\sim 2\times 10^{35}$ erg/s 
(assuming the 5 kpc distance to SS433), which is about 10\% of the soft
X-ray jet luminosity.

Using the OSA-3 software, 
we could not significantly detect  
the source in the JEM-X data.
Instead, we made use of the more detailed 
{\it RXTE} observations of 
SS433 performed simultaneously with \INTEGRAL 
in March 2004  to 
obtain the broadband 2-100 keV spectrum of SS433. The source was observed
a few days after the disk maximum opening phase. 
 The  OSA-3 software also did not allow
us to significantly detect  the source at energies above 70 keV, so 
we used the IKI software 
to obtain the 
SS433 spectrum up to 100 keV. This software has proven its
quality and efficiency in processing \INTEGRAL observations of the Galactic center
(Revnivtsev et al. 2004b).

The resulting broadband X-ray spectrum of SS433
in March 2004 is shown in Fig. \ref{fig:RXTEIBIS_sp}. It can be 
adequately fitted by the  bremsstrahlung 
emission from optically thin thermal plasma with $kT\sim 30$ keV (
model "bremss" from XSPEC package was used; see Table
\ref{sp_param}). The reduced chi-square value for $50$ dof is $\sim 0.8$, 
which corresponds to a null hypothesis probability of $\sim 0.9$. 

B. The IBIS/ISGRI count rates of SS433 in May 2003 are presented in Fig.
\ref{fig:IBIS_lc}. The X-ray eclipse at hard energies is observed to be
slightly narrower than the optical one, slightly broader than in the 
4.6-27 keV \Ginga energy range and displays extended wings 
(Fig. \ref{fig:eclips_Xop}). This is opposite to what is found in 
ordinary eclipsing X-ray binaries (like Cen X-3, Vela X-1 etc.), in which
the X-ray eclipse duration decreases with energy. 

C. The eclipse depth is observed to be at least 
80\% in hard X-rays compared to 
$\sim 50\%$ in the 4.6-27 keV band (Fig. \ref{fig:prec}-\ref{fig:eclips_Xop}).  

D. The 25-50 keV X-ray flux increases from $\sim 5$ to $\sim 20$ mCrab 
during March-May 2003 when the source precessed from the crossover 
(T2) phase to the maximum opening disk phase (T3).  
This modulation is $\sim 2$ times
larger than observed in the 2-10 keV energy band
(see Fig. \ref{fig:prec_ampl}). Thus, both precessional and 
eclipsing hard X-ray variabilities in SS433 exceed by $\sim 2$ times those  
in the standard X-ray band (Fig. \ref{fig:prec_ampl}). This suggests 
a more compact  vertical structure of the hard X-ray emitting
region in the central parts of the accretion disk.

\section{Analysis of precessional and orbital variability of SS433 in the
25-50 keV energy range}

\subsection{Precession variability}

\begin{figure}
\centering   
\epsfclipon
\epsfig{file=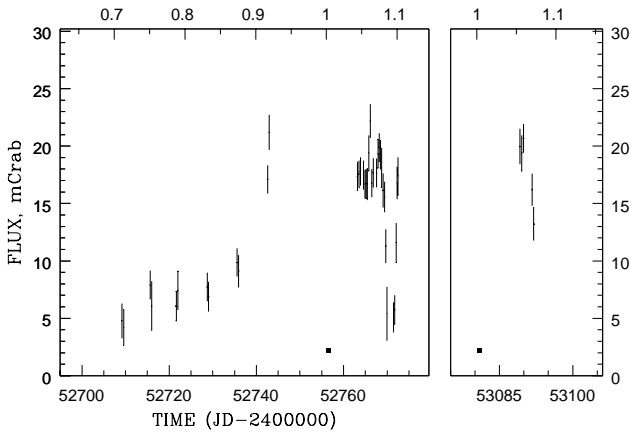,width=\linewidth}
\caption{Precessional hard X-ray (25-50 keV) 
variability of SS433. Left panel: March-May 2003.
Right panel: March 2004. The filled squares mark
the T3 precession phase according to ephemeris by Goranskij et al.
[1998ab]. 
The upper axis shows the precessional phase according to the same ephemeris.
\label{fig:prec}}
\end{figure}

\begin{figure}
\centering    
\epsfig{file=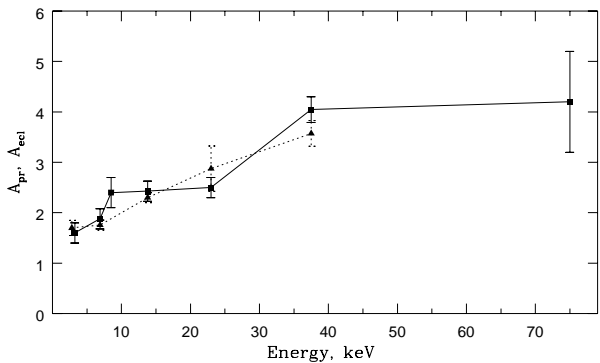,width=\linewidth}
\caption{The precession and eclipse amplitudes of SS433 at different energies in the quiet state:
 $A_{pr}$ (squares and solid line) and $A_{ecl}$ (triangles and dashed line).
\label{fig:prec_ampl}}
\end{figure}

As mentioned above, the 25-50 keV flux of SS433 varies from
$F_{min}\sim 5$ mCrab  
to $F_{max}\sim 20$ mCrab, with $F_{min}$ and
$F_{max}$ corresponding to precession phases of the disk seen edge-on
(moment T2) and at
maximum open (moment T3), respectively, yielding 
the ratio $A_{pr}\equiv F_{max}/F_{min}\sim 4$. 
In Fig. \ref{fig:prec} we present all our available \INTEGRAL 
observations of SS433. The left panel shows the 2003 data (\INTEGRAL orbits
49, 51, 53, 56, 58, 60, 67-70). The right panel shows the 2004 data (March 24-25, 
orbits 176 and 177). The uneclipsed flux of SS433 in 2004
 (when the source was observed near the T3 phase) is
comparable within the errors with the maximum flux observed near the  T3
phase in 2003 (these phases are indicated by the filled squares). This 
suggests that the precessional hard X-ray variability of SS433 stays
constant at least over several precessional cycles.

The amplitude of the precession variability in SS433 
in different X-ray bands
appears to be monotonically increasing with 
energy (Fig. \ref{fig:prec_ampl}, the squares and 
the solid line; Table \ref{tab:prec1}). 
For example, 
archive RXTE/ASM data collected over several years (only quiet states
of SS433 have been selected) indicate that $A_{pr}(1.5-5 \hbox{keV})=1.6$, 
$A_{pr}(5-12 \hbox{keV})=2.4$
(Fabrika et al. 2004). 
The errors are due to the SS433 flux at the crossover being heavily
absorbed (especially at the first crossover). The \Ginga data (Kawai et al.
1989, Yuan et al. 1995) also fit this dependence (Table \ref{tab:prec1}).
The monotonic increase of the precession amplitude with energy
is consistent with the model of an  emitting region in SS433 
as a cooling outflow. 

\begin{table}
\caption{Change of the precession amplitude of SS433 $A_{pr}$ with energy}
\label{tab:prec1}
\begin{center}
\begin{tabular}{ccccl}
\hline\hline
Energy range &   $A_{pr}$  &  Data & Ref.     \\
 keV	    &              &       &           \\ 
\hline
  1.5-5     &   1.6 & ASM/RXTE & [1] \\
4.6-9.2     &  1.88 & \Ginga & [2,3] \\
  5-12     &   2.4  & ASM/RXTE & [1]\\
9.2-18.4   &  2.43  & \Ginga & [2,3]\\
18.4-27.6   & 2.5   & \Ginga & [2,3]\\
25-50	    & 4     & \INTEGRAL & [4]\\
50-100   &  4.2     & \INTEGRAL & [4]\\
\hline 
\end{tabular}
\end{center}
\par

[1] - Fabrika et al. 2004; [2] - Kawai et al. 1989; 
[3] - Yuan et al. 1995; [4] - Cherepashchuk et al. 2003
\end{table}

For a jet of finite length or consisting
of individual evolving fragments with finite life times (Lind
and Blandford 1985; Begelman et al. 1984; Panferov and
Fabrika 1997), the intensity of radiation in the jet rest
frame is ${\cal J}_{co}=(1+z)^{2 + \alpha}{\cal J}_{obs}$,
where ${\cal
J}_{obs}$ is the observed intensity of the jet emission,  
$z=-\delta\nu/\nu$ is negative for the approaching jet and
$\alpha$ is the spectral index ($I_{\nu} \propto \nu^{-\alpha}$).
With appropriate values of the spectral index, the 
precession amplitude due to the relativistic beaming only in the case where both
jets are always seen and are not screened by the accretion disk would be
$\sim 1.1$, 
and in the case that the receding jet is fully screened by the disk it would
be $\sim 1.5$. So the large ($\sim 4$ times) precession variability found in
the hard X-ray band 25-50 keV cannot be explained only by the jet
relativistic beaming and should be 
almost fully caused by geometric screening of
the inner disk structure producing hard X-rays by the edge of the precessing
thick accretion disk. The precession light curve in the 25-50 keV energy band
exhibits a significant irregular variability which is apparently related
to the non-stationarity of the outer edges of the 
geometrically thick accretion disk.

\subsection{Hard X-ray eclipse}

The \INTEGRAL observations in May 9-11, 2003,
enable us to study in detail its hard X-ray eclipse.
We have used our data of May 9 and 11. Data for May 10, 2003, were kindly 
provided by Dr. Diana Hannikainen. The main eclipse falls on the precession
phase $\psi_{pr}=0.09$ (according to the ephemeris by Goranskij et al. 1998a),
fairly close to the T3 (disk maximum opening) phase. 

\begin{figure}
\centering    
\epsfig{file=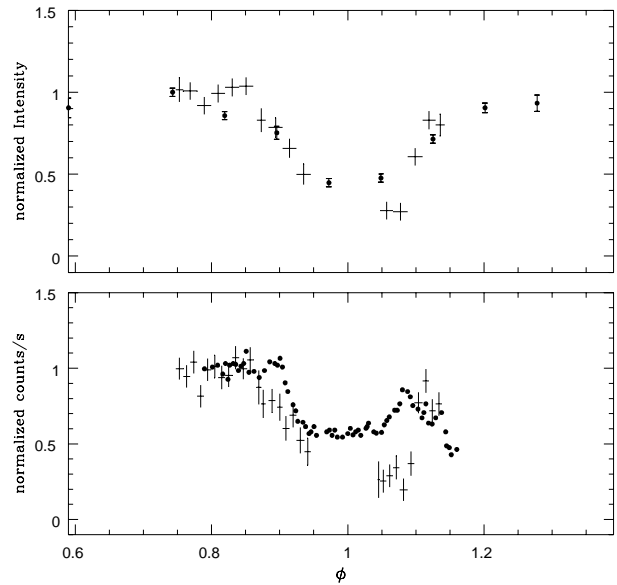,width=\linewidth}
\caption{The primary X-ray eclipse of SS433 in the 18-60 keV energy band.
IBIS/ISGRI data (May 2003). Upper panel: X-ray light curve averaged over
20000 s (10 \INTEGRAL science windows, SCW) superimposed on the
simultaneously obtained V optical photometric light curve (Crimea, SAO).
Bottom panel: the same hard X-ray eclipse light curve averaged over 10000 s
(5 SCW) superimposed on the \Ginga 4.6-27 keV eclipse (filled circles; from
Kawai et al. 1989, Yuan et al. 1995) taken at about the same precession
phase. The \INTEGRAL data is the same in both panels, but the averaging is
different.
\label{fig:eclips_Xop}}
\end{figure}

The primary X-ray eclipse in the 18-60 keV
energy band (IBIS/ISGRI data) is plotted in  Fig.~\ref{fig:eclips_Xop}. 
The points are averaged over 
5 and 10 \INTEGRAL scientific windows (10000 and 20000 s, respectively). 
For comparison, we plot the 4.6-27 keV X-ray eclipse observed by \Ginga (Kawai et
al. 1989, Yuan et al. 1995) at about the same precession phase. The upper
panel in Fig. \ref{fig:eclips_Xop} also shows the optical (V) light curve 
observed simultaneously with \INTEGRAL in Crimea and SAO (see below).
The upper panel demonstrates the absence of any noticeable
lag between optical and hard X-ray primary eclipses.  

Fig. \ref{fig:eclips_Xop} shows that 
the \INTEGRAL eclipse is up to two times deeper than that observed by
\Ginga in softer X-ray bands. 
The ascending branch of the hard 18-60 keV eclipse 
clearly shows some irregularities. These features are similar to
those observed by \Ginga in the softer 4.6-27 keV X-ray band.
If this is indeed the case, we can take 
the upper envelope of the ascending branch of the X-ray eclipse light curve
as a representative shape of the eclipse by the optical star body. 
Note that the X-ray eclipse minimum observed by \Ginga was consistent
with the synchronously observed optical one (Aslanov et al. 1993). 
This proves that there is no appreciable phase shift between the
middle of the \INTEGRAL  and \Ginga X-ray eclipses
shown in Fig. \ref{fig:eclips_Xop}.

\begin{table}
\caption{Primary X-ray eclipse depth $A_{ecl}$ variations with energy}
\begin{center}
\begin{tabular}{cccccl}
\hline\hline
Energy &  $\langle E\rangle$ &  $A_{ecl}$  &  Data & Ref.\\
 keV   &   keV               &             &       & \\ \hline

1.0-4.6     & 2.8  & 1.7 &  \Ginga    & [1,2]\\
4.6-9.2     & 6.9  & 1.8 &  \Ginga    & [1,2]\\
9.2-18.4    & 13.8 & 2.3 &  \Ginga    & [1,2]\\
18.4-27.6   & 23.0 & 2.9 &  \Ginga    & [1,2]\\
25-50       & 37.5&  3.6 & \INTEGRAL & [3]\\
\hline 
\end{tabular}
\end{center}
\label{tab:ecl1}
\par
[1] - Kawai et al. 1989; [2] - Yuan et al. 1995; [3] - Cherepashchuk et al.
2003
\end{table}

The X-ray eclipse depth as a function of energy 
is shown in Fig. \ref{fig:prec_ampl} by the dotted line. The eclipse
depth $A_{ecl}$ was defined as the ratio of the mean uneclipsed flux 
before the eclipse to the middle eclipse flux at the T3 precession phase of
SS433 in the quiet state. These data are collected in Table
\ref{tab:ecl1}. Note the tendency for $A_{ecl}$ to increase with energy, 
which may mean that the hotter parts of the X-ray emitting region 
located closer to its base
are totally eclipsed by the optical star. 

The dependence of the precession amplitude and X-ray eclipse depth on energy 
is the same, suggesting similar 
screening efficiency of the X-ray emitting region
by the accretion disk edge and by the optical star. 
This means that the size of the accretion disk can be comparable with that of
the optical star. Such a conclusion has been also inferred from 
optical photometry (Fabrika \& Irsmambetova
2002; Fabrika 2004).

\subsection{The  geometric model of the system}

\begin{figure}
\centering
\epsfig{file=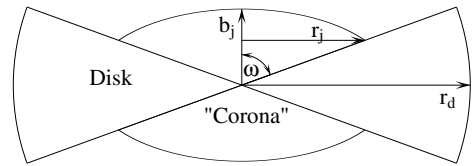,width=0.7\linewidth}
\caption{Geometrical model of the accretion disk and its ``corona''.
\label{fig:model}}
\end{figure}

To interpret the X-ray  light curve of SS433 we used a 
geometric model applied earlier
to the analysis of the \Ginga X-ray eclipse
(Antokhina et al. 1992). We consider a close 
binary system consisting of an opaque "normal" star  limited by 
the Roche equipotential surface and a relativistic object surrounded by
an optically and geometrically thick "accretion disk".  The "accretion disk" includes
the disk itself and an extended photosphere formed by the outflowing
wind. The orbit is circular, the axial rotation of the normal star
is synchronized with the orbital revolution. 

The disk is inclined with respect to the orbital plane by an angle $\theta$.
The opaque disk body (see Fig.~\ref{fig:model}) is described by the radius
$r_d$ and the opening angle $\omega$. The central object is surrounded by a
transparent homogeneously emitting spheroid with a visible radius $r_j$ and
height $b_j$, which could be interpreted as a ``corona'' or a ``jet''
(without any relativistic motion).

Here $r_j$, $b_j$ and $r_d$ are dimensionless values expressed in units of the
binary separation $a$. The radius of the normal (donor) star is determined by the relative 
Roche lobe size, i.e. by the mass ratio $q=M_x/M_v$ 
($M_x$ here denotes the mass of the relativistic object).

Only the ``corona'' is assumed to emit in the hard X-ray band, while the 
star and disk 
eclipse it during orbital and precessional motion. 
During precession the inclination of the
disk with respect to the observer changes, causing different conditions of  
``corona'' visibility. Observations of the 
 precessional variability can thus be used to obtain
a ``vertical'' scan  of the emitting structure, restricting the parameters
 $b_j$ and $\omega$.
The orbital (eclipse) variability observations scan the 
the emitting structure ``horizontally'', restricting possible values 
of  $r_d$, $\omega$, $q$ and $r_j$.
The joint analysis of precessional and eclipse variability 
enables us to 
reconstruct the spatial structure of the region in the accretion disk
center where the hard X-rays are  produced.

The position of the components of the system relative to the 
observer is determined by the binary orbit inclination angle $i=78.8^\circ$, 
the disk inclination angle to the orbital plane $\theta=20.3^\circ$, 
the precession phase $\psi_{pr}$ with $\psi_{pr}=0$ at the maximum 
disk opening of SS433 (T3, maximum separation between 
the moving emission lines)
and $\psi_{pr}=0.34, 0.66$  when the disk is seen edge-on 
(at the moving emission line crossover moments T1 and
T2, respectively).

To analyze the X-ray eclipse and precession light curve 
observed by \INTEGRAL  we have used
the following three model shapes of the hard X-ray emitting region
which can be parametrized by two parameters $r_j$ and $b_j$:
long narrow jets($r_j\ll b_j\sim a$), short narrow jets ($r_j\ll b_j\ll a$),
and short thick ``jets'', or corona ($r_j>b_j$). 
For all models the formal $\chi^2$ minimum is reached for a maximum
allowed accretion disk radius $r_d$ and the disk opening angle $\omega\sim 90^\circ$
(see discussion below).  
The accretion disk radius was limited by the distance from the relativistic
object center to the inner Lagrangian point, which exceeds the radius of the
Roche lobe around the compact star. This is natural for a supercritical
accretion disk with a strong radial outflow (Zwitter, Calvani \& D'Odorico,
1991). Further extension of the accretion disk is not justified because of
the presence of narrow absorption lines of the optical A5-A7 component in
the spectrum, which means that there is no appreciable screening of the
optical star by the disk. Such an extension of the disk would mean that the
binary system is on the common envelope evolutionary stage, and so the
narrow absorption lines from the optical star surface could not be observed.
The orbital period stability over 30 years of observations also suggests
against the presence of a common envelope in SS 433.

The results of our analysis can be summarized as 
follows.

\begin{figure}
\centering    
\epsfclipon
\epsfig{file=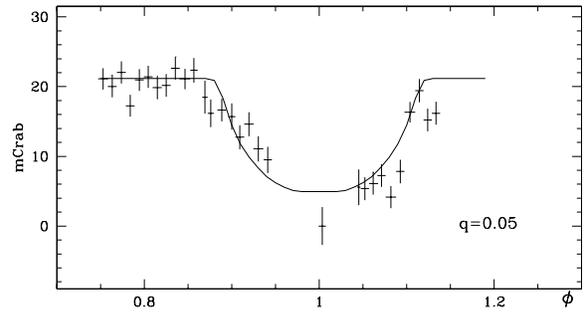,width=\linewidth}
\caption{X-ray eclipse fit with a model of narrow long X-ray emitting
jets. Precession
variability cannot be reproduced by this model although the eclipse shape is fitted.
The point at the center of the eclipse was taken from the Galactic Plane Survey scan 
in the beginning of the 70th \INTEGRAL orbit
\label{fig:jthin_l}}
\end{figure}

\begin{figure}
\centering 
\epsfclipon   
\epsfig{file=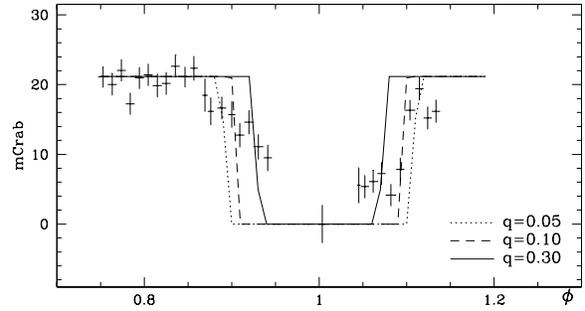,width=\linewidth}             
\caption{The same as in Fig.\ref{fig:jthin_l} but for a model of 
narrow short X-ray emitting jets for 
different values of $q$. Precession variability amplitude $A_{pr}=4$ 
is reproduced, but the eclipse shape is not.
\label{fig:jthin_s}}
\end{figure}

\begin{figure}
\centering    
\epsfclipon
\epsfig{file=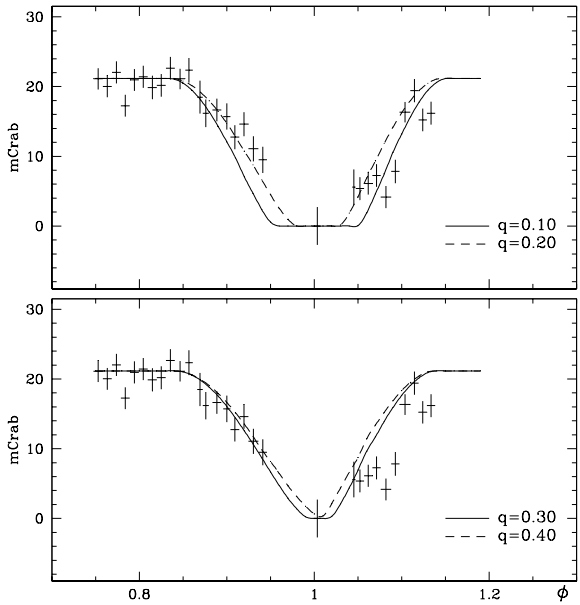,width=\linewidth}
\caption{The same as in Fig.~\ref{fig:jthin_l} but with a model of an
extended oblate X-ray emitting region (``corona'') for different $q$. The
precession variability amplitude is $A_{pr}=4$.
\label{fig:jthick}}    
\end{figure}

1. Long 
 narrow jets ($r_j \ll b_j\sim a$) outflowing from the central object
 can be only partially screened by the disk during its eclipse and precession. 
Neither the  observed eclipse depth nor the precession amplitude $A_{pr}\sim 4$ 
can be explained by this model.
(See Fig. \ref{fig:jthin_l}).

2. Short narrow jets  ($r_j\ll b_j \ll a$) 
where $b_j$ is found from the condition that the precession
variability amplitude is $A_{pr}=4$. 
The theoretical eclipse light curves for this model for 
different mass ratios $q$ are displayed in Fig. \ref{fig:jthin_s}.
For any $q$ this model is
inconsistent with the observed 18-60 keV X-ray eclipse shape.

\begin{figure}
\centering    
\epsfclipon   
\epsfig{file=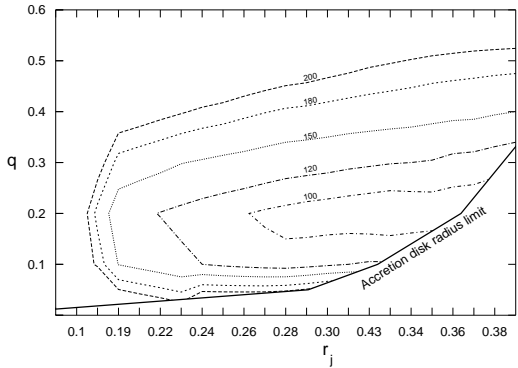,width=0.7\linewidth, angle=270}
\caption{$\chi^2$ map (26 dof) in the plane ($r_j,q$) 
for models with an extended oblate X-ray 
emitting region. The outer edge of the accretion disk 
limited by the distance to the inner Lagrangian point 
is shown by the solid line. The 
precession variability amplitude is $A_{pr}=4$. 
\label{fig:chi2}}
\end{figure}

3. An extended corona ($r_j > b_j$) gives the best fit for the 
observed hard X-ray precession amplitude ($A_{pr}\sim 4$ ) and 
the observed eclipse shape (Fig.
\ref{fig:jthick}). Minimum residuals 
in this model are attained
for the mass ratio $q=m_x/m_v \approx 0.2$ (Fig. \ref{fig:chi2}), 
which is close to that derived from
fitting the \Ginga and \ASCA observations with a narrow jet model 
(Kawai et al. 1989, Kotani et al. 1996).
If we take into
account only the upper envelope of the ascending X-ray eclipse branch (see
above), the fit with $q \approx 0.3$ seems to be acceptable as
well (Fig. \ref{fig:jthick} and  \ref{fig:chi2}).

The parameters $b_j$ and $\omega$  appear to be related 
and cannot be determined
separately. Formal $\chi^2$ decreases as $\omega\rightarrow 90^\circ$, 
without any essential
change of $q$ for $\omega\grsim 80^\circ$. 
Obviously, 
the solution with $\omega=90^\circ$
(planar disk with zero thickness)
could not provide any precessional variability, so we accepted
$\omega\sim80^\circ$. This also agrees with the estimates of the 
thickness of accretion disk from the rapid optical and X-ray variability 
corellation (Revnivtsev et al., 2004a).

From this analysis we conclude that in the framework of our 
geometrical model of the hard X-ray emitting region 
in SS433, the observed precession amplitude and X-ray eclipse shape 
are best reproduced by a broad oblate corona above an optically thick 
accretion disk. The best fit parameters are $r_j \sim 0.3 $, $b_j \sim 0.1$ and 
$\omega\sim80^\circ$, the mass ratio $q=0.2-0.3$.

Physically, such a corona could be thought of as hot rarefied
plasma filling the funnel around the jets. In this picture   
outer parts of long thin jets that are not screened by the funnel walls
are responsible for emission in soft X-rays, while
the more extended hot corona inside the funnel produces hard X-ray flux.
The estimation of the funnel size $\sim 10^{12}$~cm was obtained 
by Revnivtsev et al. (2004a) from the analysis of rapid non-coherent
variability in SS433. The opening angle of the funnel can be 
quite large, so the apparent size of the outer funnel that we
model here as an extended oblate corona over an opaque disk can be
comparable with the accretion disk size, as our best fit parameters
suggest. The nature of such a plasma inside the
funnel and how it maintains its stationarity 
requires further studies. We also note that actual detection
accuracy of the hard part of the broadband X-ray spectrum of SS433 
obtained by INTEGRAL (Fig.~\ref{fig:RXTEIBIS_sp} ) 
is insufficient to definitely rule out the
presence of the two-component structure of the X-ray emitting region
inferred from our analysis.

\section{Optical and IR photometry}

\begin{figure} 
\centering
\epsfig{file=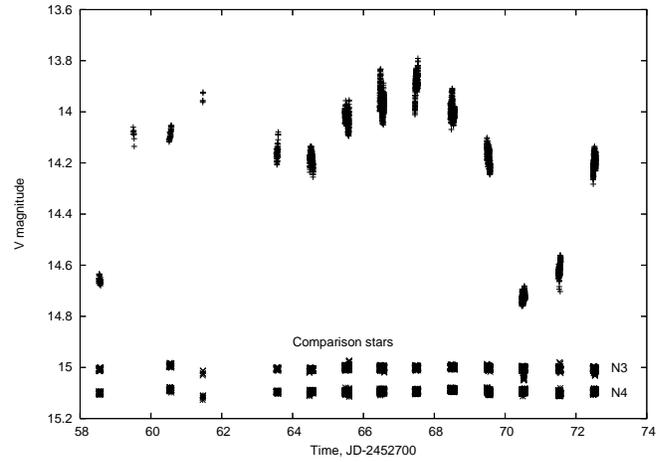,angle=270,width=\linewidth}
\caption{V-light curve of SS 433 
obtained at the RTT150 telescope (TUBITAK National Observatory, Turkey)
simultaneously with \INTEGRAL observations. Bottom: photometry
of control stars
($V_{N3}=12.93, V_{N4}=12.70$).
\label{fig:Kazan}}
\end{figure}

\begin{figure}
\centering
\epsfig{file=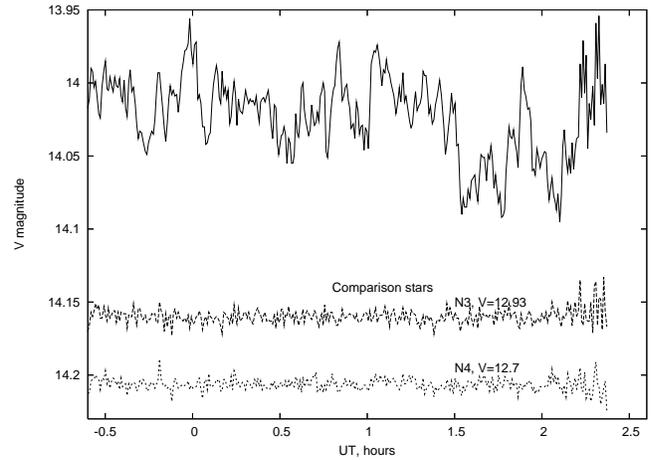,angle=270,width=\linewidth}
\caption{Rapid photometric variability of SS433 during 
the night 05/06 May 2003 
\label{fig:V050603}}
\end{figure}
   
\subsection{Optical photometry}

Photometry of SS433 was performed simultaneously with \INTEGRAL observations
by the Russian-Turkish RTT-150 telescope of Kazan University at the TUBITAK
National Observatory, Turkey. Observations were made using commercial camera
(model DW436, www.andor-tech.com) which is termoelectrically cooled to
$-60^\circ$~C. The CCD was provided for RTT150 by Max-Planck Institute for
Astrophysics (Garching, Germany). It is a low noise back-illuminated model
from EEV with 2048 x 2048 pixels of 13.5 $\mu$ size. The full field of view
is 8 x 8 arcmin with a frame reading time of 40 sec at 2 x 2 binning. To
increase the time resolution only parts of the field of view with reference
stars N 1,2,3,4,5 ( Leibowits and Mendelson, 1982) around SS433 were 
captured. The obtained V-light curve is presented in Fig.
\ref{fig:Kazan}. Strong ($\sim 0.15$ mag) intranight variability of the
source on timescales $\sim 100$ s - 100 min is clearly detected
(see Fig. \ref{fig:V050603} for the night 05/06 May 2003).

\begin{figure}
\centering
\epsfclipon
\epsfig{file=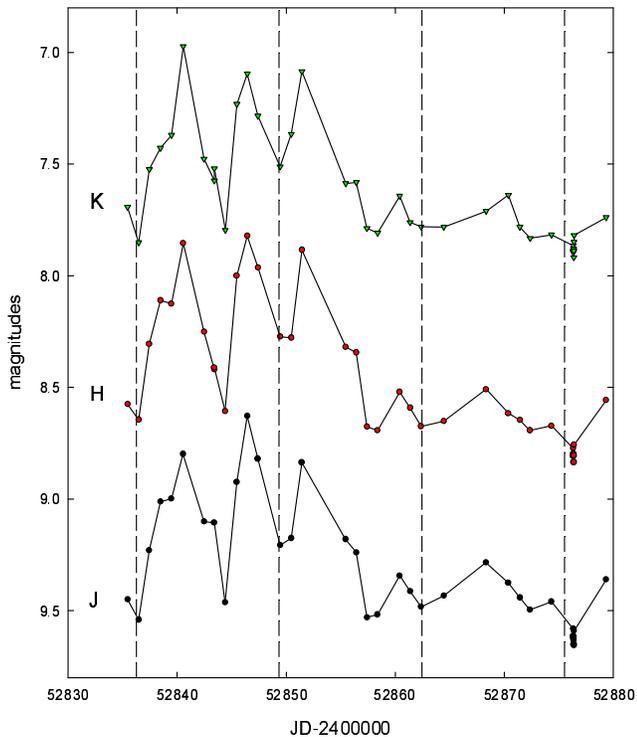,width=\linewidth}
\caption{JHK-photometrical light curve of SS433 
obtained by AZT-24 1.1-m 
IR telescope in July-August 2003 (Campo Imperatore, Italy).
The dashed vertical lines mark the primary eclipse of SS433 according
to the ephemeris by Goranskii et al. (1998a). 
\label{fig:gnedin}}
\end{figure}

The mean V light curve during the eclipse obtained at the Crimean
Laboratory of SAI (Nauchnyi, Crimea) simultaneously with the \INTEGRAL
observations is shown in the upper panel of Fig.~\ref{fig:eclips_Xop}. 
The detector was a pulse-counting,
single-channel broad-band WBVR photoelectric photometer installed on
a Zeiss-600 reflector. The photometer and software 
were designed and manufactured 
by I.I.Antokhin and V.G.Kornilov (SAI).
The data were collected in the standard way using differential
techniques. The main comparison star was C1 with magnitude $V=11.51$
(Gladyshev et al. 1980).
The typical rms measurement errors in V were 0.02-0.03.
The mean V light curve during the \INTEGRAL observations corresponds
to a quiet state of SS433.
The optical eclipse minimum 
is observed at $JD=2452770.863$, as predicted by the orbital ephemeris
given by Goranskij et al. (1998ab).

\subsection{IR-photometry}

 Near IR observations of SS 433 were obtained at AZT-24 1.1m telescope in
 Campo Imperatore (Italy) during the period of July-August 2003. Some
 observational data were obtained also in October - November 2003. The
AZT-24
 telescope at the Campo Imperatore Observatory located 2150 m  above sea
 level (a cooperation between Rome, Teramo and Pulkovo Observatories) is
used for photometric studies of variable  sources at near
infrared
 (NIR) wavelengths. The telescope is equipped with a SWIRCAM NIR 256x256
 pixels imaging camera mounted at the focal plane of AZT-24. This
 camera was built at the Infrared Laboratories in Tucson (Arizona, USA).
The SWIRCAM is equipped with a PICNIC array, an
 upgrade of the NIGMOS detector, with a working range of 0.9 - 2.5 $\mu$.
It
 yelds a scale of 1.04 arcsec per pixel resulting in a field of view of 4x4
 sq. arcmin.
        The observations were performed through standard Johnson JHK broadband
 filters. The NIR monitoring of SS433 started several orbits after the
 \INTEGRAL observations. The JHK light curves are presented in Fig.
\ref{fig:gnedin}.
 The observations were done close to the crossover precession phase, where
 the orbital modulation appears to be significantly reduced.
Coincidence of the minima in the IR and optical light curves support our 
general geometrical model.
        
\section{Optical spectroscopy}

\subsection{Spectral observations}

The optical spectroscopy of SS433 was performed at the 6-m telescope of the
Special Astrophysical Observatory
during 6 nights on April 28 and May 9-13 2003 simultaneously with the
\INTEGRAL observations at the precessional phase corresponding to the
maximum disk opening angle (T3). The Long-Slit spectrograph
(Afanasiev et al. 1995) at the telescope prime
focus equipped with a 1000x1000 Photometrics CCD-detector was used
to obtain spectra with a resolution of 3~A (1.2 A/pix).
Standard techniques were used for spectral reduction and calibration.

Table \ref{t:sp_j} lists the spectral
observations and includes date, JD of the middle of observation,
orbital  and precessional phases,
number of spectra during the night and the mean exposure,
and spectral range used. We have taken spectra in the
blue region as they are the most
informative when searching for the donor star absorption lines
(Gies et al., 2002) and include the He\,II $\lambda 4686$ emission line.
On two nights, we obtained spectra in the red region to determine the orientation
of relativistic jets and identify the moving lines  in all our spectra.
The blue spectral range was shifted redward in the
May 2003 observations because of the rising Moon.
The signal-to-noise ratio in our spectra averaged over one night at
$\lambda = 4250$\AA is $\approx 60$ per resolution element
in May 10-13 and it is $\approx 80$ on other dates. The signal-to-noise
ratio increases towards longer wavelengths.

\begin{table*}
\caption{Journal of spectral observations}
\begin{center}
\begin{tabular}{lcccccl}
\hline\hline
 Date &        JD (mid)   &   Orbital& Precession &   N&    Exposure&    Range (\AA)\\
      &        2450000+   &   phase  &   phase    &    &  (sec) &                 \\
\hline
 28.04.2003&   2758.491   &   0.054  & 0.973      & 8  &    900 &     4160--5400 \\
 09.05.2003&   2769.466   &   0.893  & 0.041      & 3  &    600 &     5170--6410     \\
 09.05.2003&   2769.507   &   0.896  & 0.041      & 4  &    900 &     3970--5220   \\
 10.05.2003&      2770.458&   0.969  & 0.047      & 4  &    600 &     5170--6410     \\
 10.05.2003&      2770.501&   0.972  & 0.047      & 4  &    900 &     3970--5220   \\
 11.05.2003&      2771.490&   0.048  & 0.053      & 8  &    900 &     4180--5430   \\
 12.05.2003&      2772.508&   0.126  & 0.060      & 4  &    900 &     4180--5430   \\
 13.05.2003&      2773.492&   0.201  & 0.066      & 4  &    900 &     4180--5430   \\
\hline 
\end{tabular}
\label{t:sp_j}
\end{center}
\end{table*}

\subsection{Search for optical star lines} 

The strongest lines in SS433, hydrogen, HeI, HeII and some FeII lines, 
appear in emission. Emission line widths in SS433 have FWHM $\sim 1000$~km/s,
corresponding to $\sim 10$~\AA\ in the blue region. The emission lines are
formed in the accretion disk wind and in gas streams in the binary system.
A detailed description of the SS433 spectrum can be found in Fabrika (2004).
In precession phases close to the crossovers (T2), hydrogen (H$\beta$ and upper
lines), HeI and FeII lines have narrow blueshifted absorption components 
with 
P\,Cyg or shell-like profiles, which are formed in gas outflow
from the accretion disk. Weaker FeII lines appear as pure absorptions in
these precession phases.
The absorption components are also strengthened in orbital
phase $\phi \approx 0.1$; they are probably formed in the disk wind
interacting with the star (Fabrika et al. 1997).

The absorption components  of the shell-like line profiles and even
the pure absorption lines are likely be formed in the disk wind too,
which follows from the strong dependence of
their intensities (Crampton \& Hutchins, 1981) and radial velocities
(Fabrika et al. 1997) on the precession phase.
Therefore, a search for the donor star spectrum in
SS433 should be made with a certain caution.  The search
should be made among the weakest
(photospheric) lines and in precession phases where the extended
disk and its outflowing wind do not intersect the line of sight.
 In addition,  stellar lines should be stronger
in the middle of the accretion disk eclipse and weaker
outside the eclipse (Gies et al. 2002).
The main criterion that the absorption lines are formed in the stellar
photosphere is a sign of the ``correct'' behavior of their
intensities and radial velocities with orbital phase.

Gies et al. (2002), Hillwig et al. (2004) detected weak absorption
lines of TiII, FeII, CrII, SrII, CaI, FeI in the blue region 4100--4600
\AA\ of the spectrum. They observed SS433 near the primary eclipse and
in precession phases where the disk is the most "face-on".
These lines are believed to belong to the donor
star of SS433 with an A-type spectrum.

One may also try to search for stellar lines in the yellow and red regions
among the weakest metallic lines, however these regions are crowded with
strong moving lines of hydrogen and HeI formed in the relativistic jets. The
moving lines are extremely variable and have structured profiles. This
complicates the search for weak lines and the study of their behavior from
date to date. For this reason we have investigated only the bluer parts of
our spectra. The most informative region for stellar line studies is
4200--4340~\AA\, where no strong emission lines and no appreciable moving
lines appear at these precession phases.

The 4100--5300~\AA\ SS433 spectra have been compared with
those of the known galactic supergiants with different temperatures.
We used publicly available
(http://webast.ast.obs-mip.fr/stelib/) library of stellar spectra
(STELIB, Le Borgne et al. 2003). These spectra were taken
with the same spectral resolution (3~\AA) as our spectra of SS433.
For identification of lines we have used spectral atlases of Deneb (A2 Iae)
(Albayrak et al. 2003) and o Pegasi (A1 IV) (Gulliver et al. 2004).
Spectra of SS433 summed over one night were used for the line
identification.

\begin{figure*}
\centering
\epsfig{file=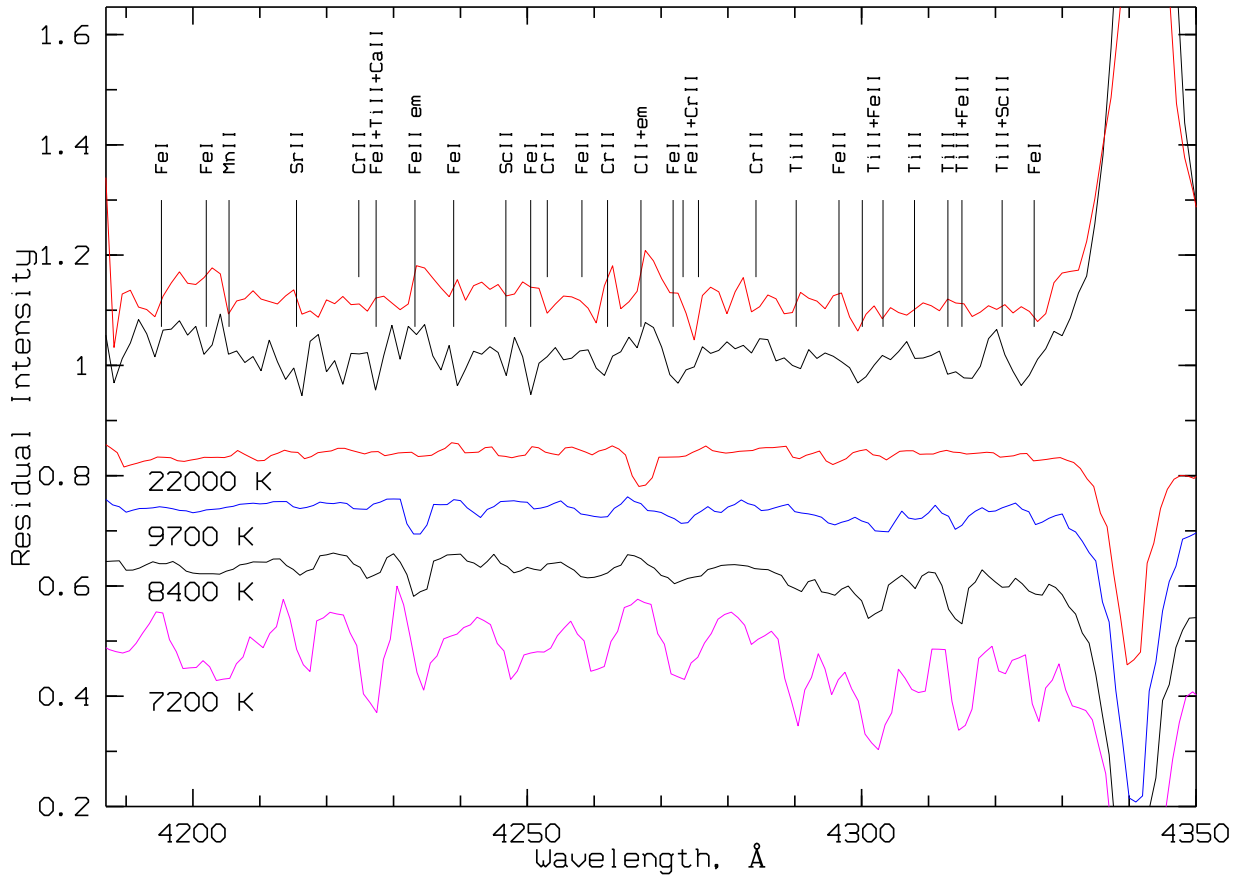,width=0.7\linewidth,angle=-90}
\caption{
Spectra of SS433 and spectra of  four supergiants of known
temperatures (Le Borgne et al., 2003) for a comparison.
 The top spectrum was taken outside the eclipse and 
averaged over nights
09.05.2003 and 12.05.2003 (evenly spaced from the eclipse center
orbital phases 0.89 and 0.12). The second from the top spectrum
was taken inside the eclipse and averaged over nights 28.04.2003 and
11.05.2003 (the orbital phase 0.05).}
\label{fig:SPklass}
\end{figure*}

\section{Spectral type of the optical star}

 Fig. \ref{fig:SPklass} shows spectra of four supergiants
and two SS433 spectra taken outside and inside the eclipse,
each averaged over two nights. The eclipse-out spectrum was
averaged over nights 09.05.2003 and 12.05.2003 corresponding
to orbital phases 0.89 and 0.12 
evenly spaced from the eclipse center 
(the top spectrum). The in-eclipse spectrum was averaged over
nights 28.04.2003 and 11.05.2003 corresponding to orbital phase 0.05
(spectrum second from the top in the figure).
The supergiants shown in Fig. \ref{fig:SPklass} are HD36673 (F0I, 7,200~K),
HD39866 (A2I, 8,400~K), HD87737 (A0I, 9,700~K) and HD164353 (B5I, 22,000~K).
All spectra were normalized and shifted along the vertical axis for
better visualization.

With our spectral resolution,
practically all absorption lines appear as complex blends.
In Fig.~\ref{fig:SPklass} the lines observed
out of the eclipse are marked with short vertical bars;
those seen
in the eclipse or present in both spectra are marked with long bars.
One can see that the strongest absorptions in the spectra of
supergiants are also present in SS433 and these absorption
lines are deeper in the spectrum during the primary minimum.
FeII$\lambda4233$ emission is seen (FWHM$ \sim 10 $~\AA),
CII$\lambda 4267$ and NIII$\lambda 4196,4200$ + HeII$\lambda 4200$
emissions are also marginally present.

The intensity of absorption lines during the eclipse allows us to
estimate the effective temperature of the
 donor star to be $T<9000$~K. This limitation was obtained from the fact
that the absorption lines in SS433 cannot be deeper than those in spectra of
supergiants, because the accretion disk is not totally eclipsed in the
primary minima. The relative intensities of the strongest metallic
absorption lines indicate a temperature of $T=8000\pm 500$ K, implying that
the optical spectral class of the companion is A5-A7I.

\subsection{Evidence for a heating effect in the illuminated hemisphere}

The analysis of different absorption lines of the optical star
reveals a strong heating effect in the illuminated star's atmosphere.
During the disk eclipse egress low-excitation
absorption lines strongly weaken. The stellar hemisphere illuminated by
the bright accretion disk in SS433 probably has a temperature of $\sim 20,000$ K,
as the presence of CII$\lambda 4267$ absorption + emission line
suggests (the top spectrum in
Fig.~\ref{fig:SPklass}). This absorption line is the strongest one in
this spectral region among supergiants with temperatures T$ > 15,000$~K.
The evolution of the blend $\lambda4273$ (FeI$\lambda 4271.7$ +
FeII$\lambda 4273.3$ + CrII$\lambda 4275.5$) is seen: in the
eclipse  center the low-excitation FeI line is stronger,
while out of the eclipse FeII+CrII lines are enhanced.
Other FeI lines also appear only in the eclipse.

\begin{figure*}
\centering
\epsfig{file=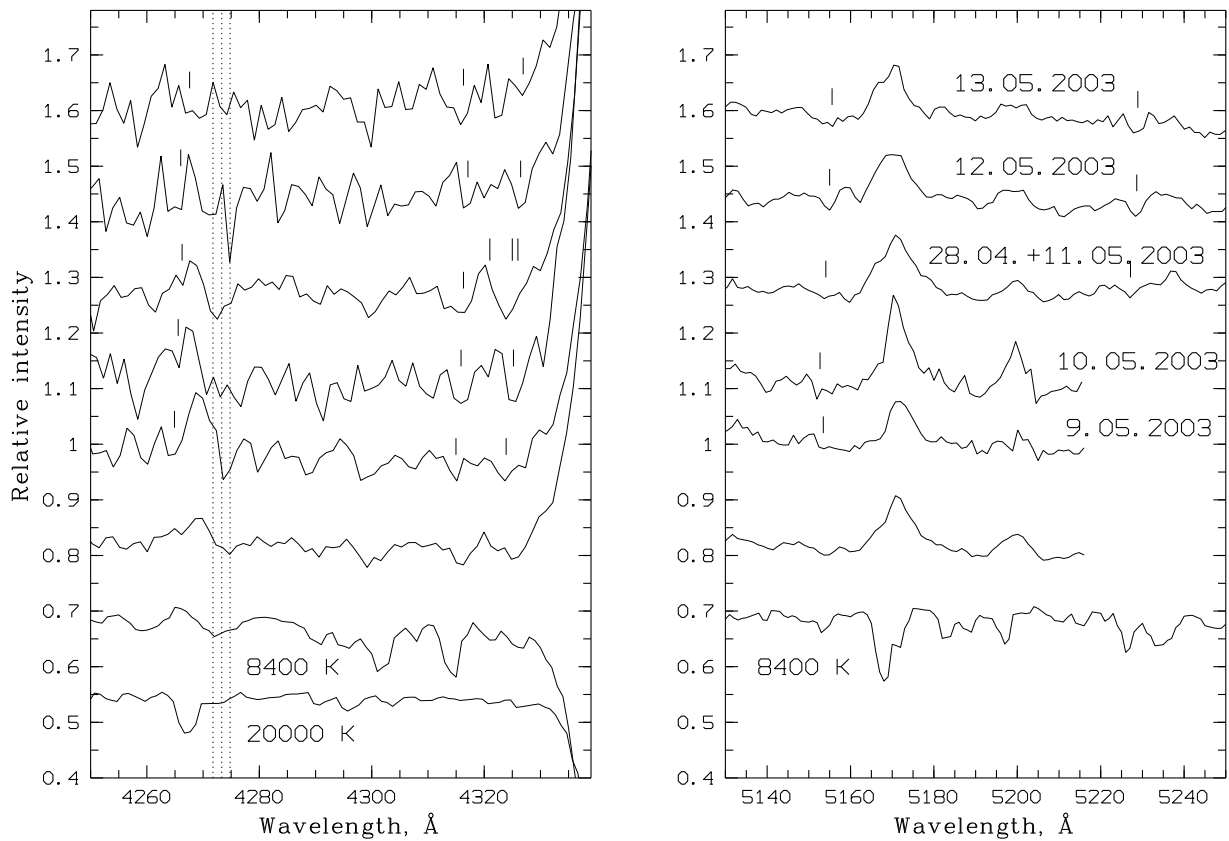,width=0.9\linewidth}
\caption{Evolution of absorption lines in the
optical spectra of SS433 taken at different orbital phases
(see Table \ref{t:sp_j}).
For comparison, spectra of two supergiants (8,400 K and 22,000 K) are shown
in the bottom. Radial velocity of the optical star is
traced by different absorption lines (vertical bars).
 On the left panel, the 
third from bottom is the SS433 spectrum averaged over all nights
with taking into account the radial velocity shifts from Fig. \ref{fig:radvel}.
See text for more details.}
\label{fig:demo21}
\end{figure*}

These effects are illustrated by Fig.~\ref{fig:demo21}, in which spectra
of the standard stars with effective temperatures 8,400~K and 22,000~K
are shown together with spectra of SS433 obtained on 9.05, 10.05, 28.04+11.05,
12.05 and 13.05 (orbital eclipses fell on nights 10.05 and 28.04+11.05).
The orbital phase rises from bottom to top in the figure. There are
emission lines in these spectral fragments~--- the broad structured emission
of CII$\lambda 4267$, and stronger emissions FeII$\lambda 5169$+MgI$\lambda
5167, 5173$ and FeII$\lambda 5197$. Note that emission lines,
which presumably are formed
in the disk wind, should move blueward
(in phase with the compact object) in the orbital phases of our
observations. In contrast,
absorption lines, which are formed in the donor photosphere, must move
 redward during our observations.
The strength of the CII$\lambda 4267$ absorption component (marked
by vertical bars in Fig.~\ref{fig:demo21}) increases notably out of
the eclipse. The
 complex
blend $\lambda4273$ (FeI+FeII+CrII) is marked by the dotted
lines in the figure.

Fig.~\ref{fig:demo21} illustrates absorption lines (marked by the vertical
bars)~--- the blend $\lambda 4314$ of TiII$\lambda
4313$+FeII$\lambda 4314$+TiII$\lambda 4315$;
ScII$\lambda 4320$ line and the blend
ScII$\lambda 4325$ + FeI$\lambda 4326, 4327$. However the last blend
may be distorted by the strong blue wing of H$\gamma$ in the middle of
eclipse. Other absorption lines shown in Fig.~\ref{fig:demo21}
(right panel) are $\lambda 5154$ (TiII+FeII) and  $\lambda 5227$ (TiII+FeI).
It is seen that these lines shift redward with time.

 In Fig.~\ref{fig:demo21} we also show the spectrum of SS433
averaged over all nights of observations (the third from bottom).
When averaging spectra obtained on individual nights we  shifted them
to zero velocity according to the radial velocity curve from
Fig. \ref{fig:radvel}. The average spectrum has a better signal-to-noise
ratio.
All the absorpton lines remain in the spectrum and they are similar
to the corresponding lines in the spectrum of the standard supergiant (8,400~K).
The shallow emission line (presumably CII$\lambda 4267$) has an absorption
component.
In our recent spectral observations of SS433, which were
carried out around primary
mimima on August 24, 2004 and September 6, 2004 (the precession phase is
the same, i.e. the maximum disk opening) we confirmed the presence of the absorption lines
studied in this paper. Both the spectral resolution and signal-to-noise
ratio are better in the new observations. Analysis of the new data will be
published elsewhere.

\subsection{Measurements of the radial velocity of the optical star}

\begin{figure}
\centering
\epsfig{file=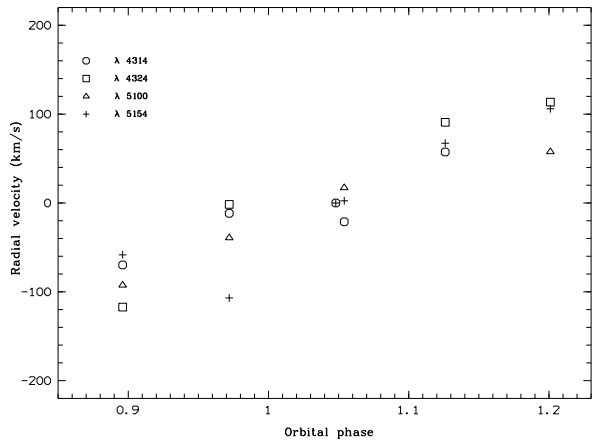,width=0.7\linewidth,angle=-90}
\caption{Radial velocity curve of the optical companion of SS433
obtained from 4 individual absorption lines over 6 nights.
\label{fig:rvindid}}
\end{figure}

It is important to stress that the strongest absorptions (FeII, HeI)
are disfavored for the optical star radial velocity analysis, as they
most probably formed in the powerful outflowing disk wind (Fabrika et al.
1997). We studied radial velocities of weak absorption lines of metals.
It is easy to confuse the line identification when one traces a line
from date to date because of the spectral variability and orbital
motion  of the companions.
In Fig. \ref{fig:rvindid} we present individual
absorption line radial velocities for four best-suited lines
averaged over one night, as a function of the orbital phase for 6 nights.

\begin{figure}
\centering
\epsfig{file=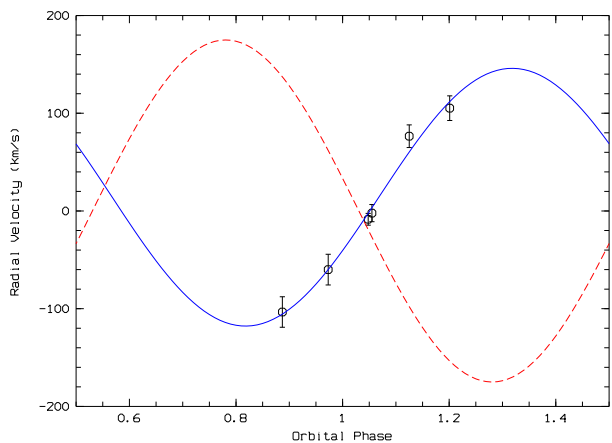,width=0.7\linewidth,angle=-90}
\caption{Mean radial velocity curve of the optical companion of SS433
measured from 22 individual absorption lines. The accretion disk
radial velocities as measured by HeII$\lambda 4686$ emission
(Fabrika and Bychkova, 1990) is shown by the dotted line.
\label{fig:radvel}}
\end{figure}

The radial velocity curve measured by the most reliable collection of
22 absorption lines in the spectral range 4200-5300~\AA\  is shown in
Fig. \ref{fig:radvel}. Note that with 
our spectral resolution the absorption lines in the SS433 spectrum
are seen as blends containing 2-3 lines,
and we measured their relative radial velocities because of unknown
"laboratory" wavelengths of the blends.
Relative intensities of the lines change from date to date
through the eclipse.
For this reason we measured only those lines and on only those dates
where the lines were the most convincingly detected. The radial velocity
 curve (Fig. \ref{fig:radvel}) has been obtained by coadding
radial velocity curves of individual lines, and the zero radial velocity
of all lines on 11.05.2003 (the orbital phase 0.048) was adopted.
If a line was not detected on 11.05.2003, the zero radial velocity
was assumed for 28.04.2003 as well (the orbital phase 0.054). If the measured
radial velocity on 11.05.2003 disagreed with the whole radial
velocity curve of a given line, an additional shift
 for the curve was applied. However no shifts in excess
of 20~km/s were done. This is the reason why the radial
velocity error in  11.05.2003 is
small, but not zero.

Thus, the radial velocity curve for absorption lines in Fig. \ref{fig:radvel}
consists of individual radial velocity curves. Each point of the
curve includes from 10 to 17 individual measurements. Most
measurements were carried out at  the primary minimum as there the
absorption lines are deeper.

The derived radial velocity semi-amplitude of the donor star is
$K_v=132\pm 9$~km/s, the gamma-velocity of the binary system is
$v_\gamma=14$~km/s with a formal fit uncertainty of $2$~km/s.
The absorption line radial velocity transition through
the $\gamma$-velocity occurs at the middle of the optical eclipse
($\phi_b=0.07$), confirming that the lines actually
belong to the donor star. Note that in the gamma-velocity $v_\gamma$
and in the transition phase $\phi_b$  some systematic errors
can be present
 because of the method used. However they are less
than 30~km/s (no bigger shifts were applied to radial velocity
curves of individual lines) and 0.05 correspondingly.
The main observational bias in studying radial velocities of the
donor star is the faintness of  absorption lines outside the primary
minima and strong intrinsic spectral variability of SS433. To
confirm the derived radial velocity curve further spectral observations
are needed.

Our results confirm the earlier determination of $K_v$ by Gies et al.
(2002) which was carried out also at the maximum disk 
opening phases. Note that
spectroscopic observations by Charles et al. (2004) were performed at the
crossover phase of SS433 when the accretion disk is seen edge-on. Such a
phase is disfavored for the donor star
radial velocity analysis as strong gas outflows contaminate
the disk plane; selective absorption in this gas affects
the true radial velocity of the donor star.

The heating effect of the donor star also distorts the radial
velocity semi-amplitude. The analysis (Wade and Horne 1988,
Antokhina et al. 2005) indicates that the true value of the
semi-amplitude of the radial velocity curve as derived from these
absorption lines can be reduced to
 $\sim 85$ km/s (see below).

Recently Hillwig et al. (2004) obtained an estimate of the radial velocity
amplitude of the donor star $K_v=45 \pm 6$~km/s and the gamma-velocity
$v_\gamma=65\pm 3$~km/s.
In our opinion, some strong lines (which actually are shell-like
lines on the background of faint emission) could contribute to the radial
velocity cross-correlations of Hillwig et al. (2004). For example,
the two strongest lines in  their Fig.\,5 FeII$\lambda 4550$
and FeII$\lambda 4584$ are in emission in our spectra in the middle of
eclipse.

The KPNO-2003 observations of Hillwig et al. (2004) were made in the orbital
phases 0.85-1.07. Absorption lines formed in the 
rotating and extended envelope of the
 donor could be probably observed.
In these
 orbital phases, the side of the envelope
projected onto the strong continuum source
 (the accretion disk) moves away from the observer.
Then the amplitude of the line shift can be expected to
be $\sim 1/2 V_{equat} \sim 60$ km/s.
This
 might explain the positive shift of the system velocity
+65~km/s found by those authors.

\begin{figure}
\centering    
\epsfig{file=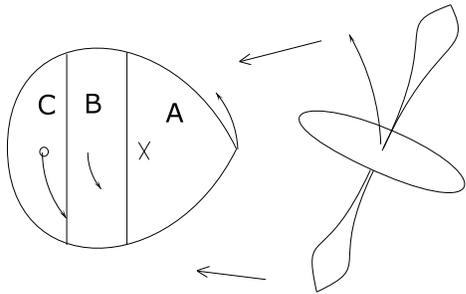,width=0.7\linewidth}
\caption{The spectroscopic center of the normal star (circle) heated by its
companion is shifted with respect to the gravity center (cross). The arc
arrows indicate the orbital motion. Zone "A" is the most effectively
illuminated by the compact source, disk and jet. Zone "B" is also
illuminated by outer parts of the disk and jet and by scattered radiation.
The low excitation potential absorption lines used for radial velocity
measurements can be formed in zone "C" which is not heated. Due to this
effect, the spectroscopically derived radial velocity of the optical star is
overestimated.
\label{fig:illum}}
\end{figure}

\section{Implications for binary masses}

Comparison of radial velocities of the accretion disk ($K_x=175$ km/s, 
Fabrika and Bychkova 1990) and optical star ($K_v=132$ km/s) yields the
mass ratio in the SS433 system $q=m_x/m_v=K_v/K_x=0.75$ (here 
$m_x$ and $m_v$ stand for relativistic object and optical star mass,
respectively). 
Taken at face value, these $q, K_v$ values would
lead to the optical star
mass function $f_v\approx 3.12 M_\odot$ and the binary component 
masses $m_x=18 M_\odot$, $m_v=24 M_\odot$. However, such a large mass ratio
is in a strong disagreement with the observed duration of the X-ray 
eclipse, which suggests a much smaller mass ratio $q\sim 0.2-0.3$. 
We stress that the binary inclination angle in SS433 ($i=78^o.8$) 
is fixed from the analysis of moving emission lines. 

There are two possibilities: (1) either the model we used to fit  X-ray 
eclipses should be modified, or (2) the value of $K_x$ and $K_v$ 
are influenced by additional physical effects. 
Although there are some reasons to modify the model (e.g., 
the asymmetric shape of the hard X-ray eclipse, 
which may suggest an asymmetric wind outflow
from the illuminated part of the optical star and wind-wind collision),
 we shall consider here only 
the second possibility. The reason is that 
the actual value of $K_v$ should be 
decreased in order to account for the observed heating effect.

Let us assume the mass ratio in the system to be $q=0.3$, as 
with this value we can satisfactorily describe the width of X-ray eclipse
and the observed X-ray precession amplitude
(see Fig. \ref{fig:jthick}). Taking $K_v=132$ km/s yields
$f_v\approx 3.1 M_\odot$ and $m_x=62 M_\odot$, $m_v=206 M_\odot$, 
$K_x=440$ km/s. Clearly, this is an unacceptable model. 
Now let us decrease $K_v$ down to 85 km/s, the lower limit that follows
from  a more accurate treatment of the heating effect in the radial velocity
curve analysis (Wade and Horne 1988, Antokhina et al. 2005). This
would yield a better fit with $m_x=17 M_\odot$, $m_v=55 M_\odot$, and
$K_x=283$ km/s, still too high to be acceptable. 
Less than half of the stellar surface is sufficiently cool to
give the absorption lines under study (e.g. due to sideway heating from
scattered UV radiation in the strong accretion disk wind).
This additionally decreases the value of the actual radial velocity
semi-amplitude. Fig. \ref{fig:illum} illustrates these considerations.

So taking, for example, $K_v=70$ km/s and $q=0.3$ yields the
optical star mass function $f_v=0.46 M_\odot$, binary masses
$m_x=9 M_\odot$, $m_v=30 M_\odot$ and the optical star radius $R_v\sim 40
R_\odot$. This radius is compatible with a typical bolometric  
luminosity of a $30 M_\odot$ A5-A7 supergiant with $T_{eff}\sim 8500$~K.
In this solution, $K_x\simeq 233$ km/s, larger than the measured value 175
km/s, but the true value of $K_x$ may be affected by the strong accretion
disk wind.

\section{Discussion}

The interpretation of the binary parameters presented above is 
strongly based on the binary mass ratio $q=0.3$ as inferred from 
modeling the \INTEGRAL observations of the hard X-ray eclipse
with account of the observed amplitude ($\sim 4$) of the 
hard X-ray precessional variability of the system.
The model we use (an optical star filling its
Roche lobe + thick accretion disk with a hot corona) is simplistic and
cannot perfectly reproduce the apparent asymmetric shape of the
X-ray eclipse. In the real situation, an additional X-ray absorption 
by outflowing gaseous streams, extended stellar envelope or wind 
seems quite plausible. If we interpret 
the complex shape of the hard X-ray egress as being due to additional
variable absorption effects, we obtain the geometry of the hard X-ray 
emitting region in the form of an extended oblate corona over the
accretion disk.
However, a straightforward 
interpretation of the joint RXTE+\INTEGRAL spectrum of the source
observed in March 2004 is also possible by a single 
thermal emission with a temperature of $\sim 30$~keV. No
additional  emitting region is required to fit 
this broadband X-ray spectrum.
 
There are two possible solutions. First, 
we could try to describe the observed hard X-ray spectrum by 
a more complex model involving two components, a hot corona + 
cooler thin jet radiating in the standard X-ray band.
Second, we could try to accept the thin cooling jet interpretation 
of the X-ray spectrum, but then would have to admit a different shape of
the X-ray eclipse. Our modeling shows that 
a thin short jet could reproduce the observed 
amplitude ($\sim 4$) of hard X-ray precessional variability 
in SS433. In that case, the shape of the X-ray eclipse must be more
sharp, with shorter ingress/egress times than actually observed 
(cf. Fig.~\ref{fig:jthin_s}). 
If we look at the data, such a shape can indeed be
found at the egress phase of the eclipse (see Fig. \ref{fig:eclips_Xop}). 
The (apparently longer) ingress time could be due to additional
(apart from the opaque star body)  absorption of hard X-ray emission by an optically
thick gaseous stream outflowing from the optical star. 
Apparently the same smooth ingress and egress shape of the eclipse 
in softer X-rays (the \Ginga and \ASCA data) in that case  
could be due to the 
soft X-ray emitting region located further upstream of the
jets. This interpretation also explains the shallower soft X-ray eclipse. 

Clearly, in this situation we need more observations of the 
hard X-ray eclipse to confirm 
the asymmetric ingress/egress eclipse shape and flatness of its
bottom. We hope to do this in 
future \INTEGRAL observations of SS433.   

\section{Conclusions}

The results of the 
\INTEGRAL observations of the peculiar binary system SS433 
in 2003-2004 can be summarized as follows.

1. Hard X-ray emission up to $\sim 100$ keV was discovered from this
superaccreting microquasar with uneclipsed hard X-ray luminosity close to
the maximum opening precession phase
$L_x(18-60 \hbox{keV})\sim 4\times 10^{35}$
erg/s, $L_x(60-120 \hbox{keV})\sim 2\times 10^{35}$ erg/s
(assuming the 5 kpc distance to SS433), which is about 10\% of the soft
X-ray jet luminosity.

2. Persistent precessional variability in hard X-rays was discovered with
the maximum to minimum flux ratio $\sim 4$, which is twice as large as in
the softer X-ray band.

3. The observed hard X-ray orbital eclipse is found to be
in phase with the optical and near infrared eclipses.  
The hard X-ray eclipse is observed to be at least two times
deeper than the soft X-ray eclipse. The width of the X-ray eclipse increases 
with energy, which is opposite to what is observed in classical X-ray binary
systems. 

4. The broadband X-ray spectrum 2-100 keV of SS433 simultaneously 
obtained by \INTEGRAL and RXTE in March 2004 when the source was in its
flaring state can be fitted by bremsstrahlung emission from optically
thin plasma with a temperature $kT\approx 30$ keV. 

5. Optical spectroscopic observations of SS433 on the SAO 6-m telescope
which were performed in the framework of the multiwavelength \INTEGRAL
campaign of the source confirmed the spectral class of the optical star 
to be A5-A7I. The radial velocity curve of the optical component
was obtained with a semi-amplitude $K_v=132\pm 9$ km/s near the maximum
disk opening precession phase (T3). The spectroscopic data revealed a 
heating effect of the hemisphere of the optical star illuminated by
the accretion disk. 

6. From the analysis of the hard X-ray eclipse and precession variability 
in SS433 we estimate the mass ratio in this binary system to be 
$q\sim 0.3$. This mass ratio and 
radial velocity measurements corrected 
for the heating effect enable us to 
evaluate the masses of the optical and compact star in SS433 to be 
$M_v\approx 30 M_\odot$, $M_x \approx 9 M_\odot$, respectively. This  finding 
lends further support to the presence of a black hole in this 
peculiar superaccreting galactic microquasar.

\begin{acknowledgements}

The authors thank E.M.Churazov for developing methods and analysis
of the IBIS data and software. We acknowledge
Vitaly Goranskij for comments and discussion. We also
ackhowledge E.K. Sheffer, S.A. Trushkin, V.M. Lyuty, S.Yu. Shugarov, N.A.
Katysheva, A.A. Lutovinov for discussions and collaboration. We especially
thank Dr. D. Hannikainen for providing us with data on SS433 X-ray egress
observations and M.G. Revnivtsev for processing {\it RXTE} spectra of SS433.
IFB thanks the Turkish National Observatory night assistants Kadir Uluc and
Murat Parmaksizoglu for their support in the photometric SS433 observations.
A.N. Burenkov is acknowledged for help in spectral observations. The work
of SNF is partially supported by the RFBR grant 04-02-16349. The work of
KAP, ANT and IEP is partially supported by the RFBR grant 04-02-16720.
The work of KAP was also partially supported by the Academy of
Finland through grant 100488. 
NIS and IEP acknowledge the financial support from the RFBR by 
the grant 03-02-16068.
SVM and ES acknowledge the European Space Agency for support and the Integral Science Data Center
(Versoix, Switzerland) for providing computing facilities.
ANT also acknowledges the financial support from the Russian Federation
President Grant Program through grant MK-895.2003.02. The work of AMCh, EAA,
EVS and ANT was partially supported through the grant of Leading Scientific
Schools of Russia NSh-388.2003.2 and RFBR grant 02-02-17524. NAS and IFB
acknowledge the support of RFBR grant 02-02-17174 and grant of Leading
Scientific Schools of Russia NSh-1789.2003.2. EAB was partially supported by
the RFBR grant 03-02-16133. The IR observations (YNG and AAA) are partially
supported by the RFBR grant 03-02-17223a , the Program of Prezidium RAN
"Nonstationary Processes in Astronomy" and the FBNTP "Astronomy".
We are grateful to the anonymous referee for the useful remarks which helped
us to improve and clarify the article.

\end{acknowledgements}

\end{document}